# Neutron diffraction studies and the magnetism of an ordered perovskite: $Ba_2CoTeO_6$.

**S. A. Ivanov**,
Karpov' Institute of Physical Chemistry, Moscow, Russia
**P. Nordblad, R. Mathieu**,
Department of Engineering Sciences, The Ångström laboratory,
Uppsala University, Sweden
**R. Tellgren\***,
Department of Materials Chemistry, The Ångström Laboratory, Uppsala University, Sweden
**C. Ritter**,
Institute Laue Langevin, Grenoble, France

**Abstract**

The complex perovskite $Ba_2CoTeO_6$ (BCTO) has been synthesised, and the crystal structure and magnetic properties have been investigated using a combination of X-ray and neutron powder diffraction, electron microscopy and dielectric, calorimetric and magnetic measurements. It was shown that at room temperature this compound adopts the 6L-trigonal perovskite structure, space group P-3m (N 164) (a= 5.7996(1)Å, c= 14.2658(3)Å). The structure comprises dimers of face-sharing octahedra as well as octahedra which share only vertices with their neighbours. A long-range antiferromagnetically ordered state has been identified from neutron diffraction and magnetic studies. The magnetic diffraction peaks were registered below the magnetic transition at about 15 K and a possible model for the magnetic structure is proposed. The structural and magnetic features of this compound are discussed and compared with those of other Co-based quaternary oxides adopting the perovskite structure.

## 1. Introduction

Multiferroics have become an object of growing interest due to the coexistence of magnetic order and ferroelectric polarization in a single-phase material [1–7]. The interest is motivated not only by the rich physics connected with the interaction of electric and magnetic properties of a crystal, that is expected in this class of solids, but also by their technological potential. Indeed, multiferroic materials with significant magnetoelectric (ME) coupling open promising opportunities for designing novel microelectronic and spintronic devices [7–9]. However, most magnetic ferroelectrics tend to have low magnetic ordering temperatures and are often antiferromagnets, in which the magnetoelectric effect is intrinsically small [9]. Generally, ferroelectricity is not compatible with ferromagnetism and therefore the chances to find new magnetoelectric materials are low but feasible [4]. Reports of large ferroelectric polarization in thin films of BiFeO$_3$ [10] and strong magnetoelectric coupling in TbMnO$_3$ [11] stimulated enhanced activity in the field of multiferroics. Recent observations of magnetoelectric effects in materials with complex magnetic structures, in which the ferroelectricity is induced by the magnetic order, also opened new prospects in the quest for new multiferroics [12]. Perovskites represent one of the most important classes of functional materials and have also played a leading role in multiferroic research [3].

\*Corresponding author

Search for magnetoelectric materials showing ferromagnetism and ferroelectricity simultaneously, in the same phase, begun in the 1950s with the substitution of diamagnetic cations by paramagnetic ones in the perovskite B-sublattice [13-17]. At the same time it was found that the presence of diamagnetic Te$^{6+}$ cations could be useful in stabilising ferroelectric properties in metal oxides with perovskite structure [15]. When the B-sites in the perovskite are partly occupied by transition metal cations, the magnetic properties are strongly influenced by the ordering of the B-site cations. Until now only few compounds have been reported to exhibit both ferroelectric and ferromagnetic properties in the same phase [9,16-20], and only data on their ferroelectric, T$_{CE}$, and ferromagnetic, T$_{CM}$, transition temperatures are available but no detailed information on their crystal and magnetic structures.

## 2. Motivation

In our recent work on multiferroics we have focused the interest on the dielectric and magnetic properties of Fe-based perovskite-type compounds [21-23]. However, also Co-based perovskites have promising magnetic and electric properties. The richness of the physical properties of the cobaltites is related to the ability of the Co cations to adopt not only several oxidation states but also various spin states. The structure of Ba$_2$CoTeO$_6$ (BCTO) is related to the hexagonal 6L-BaTiO$_3$ structure. Hexagonal perovskites, in which octahedra share faces and form chains along the hexagonal *c* axis, are less common than cubic perovskites, where all octahedra share corners, and typically form when the A cations are too large to be accommodated in the A site of the BO$_6$ framework. BCTO was initially synthesised and investigated in the mid sixties [24, 25] and then forgotten for a number of years. However, the compound experienced a regained interest [26–31] from the prospect of finding perovskite oxides with high Curie temperature. It was then also found, that an antiferroelectric phase transition, accompanied by anomalous dielectric properties occurs in BCTO close to room temperature [31].

There is a controversy concerning the crystal structure of BCTO, particularly with respect to the true space group [24, 25, 29]. According to Bayer [24], the three tellurates Ba$_2$BTeO$_6$, with B = Ni, Co and Zn have the same structure described as the hexagonal 12- layer structure. But in the study of these tellurates by optical spectroscopy [26-28] it was found that the Co compound has another structure: the c axis of the hexagonal cell is halved as compared to the nickel and zinc compounds. The 12-layer structure of Ba$_2$NiTeO$_6$, becomes a 6-layer one for Ba$_2$CoTeO$_6$ suggesting a complete ordering in the distribution of the Co and Te cations on the *1a* and *2d* sites [27, 28].

The structural properties of BCTO have not yet been studied in detail, and the role of the structural change on the magnetic and dielectric properties is not clear. The precise knowledge of the nuclear and magnetic structures of this magnetoelectric has a paramount importance in the interpretation of the ferroic and magnetic properties. An understanding of how the chemical environment affects the coexistence of ferroic and magnetic ordering is the main aim of this investigation. In this paper we add to previous studies a high-resolution NPD investigation of the thermal evolution of the crystal structure and the long-range magnetic ordering of BCTO. The study is complemented with magnetic and calorimetric measurements, which are discussed in the light of subtle structural peculiarities.

**3. Experimental**

3.1. Sample preparation
A high quality polycrystalline sample of BCTO was prepared by a conventional solid-state sintering procedure. $BaCO_3$, $CoCO_3$ and $TeO_3$ were used as starting materials. Appropriate amounts with respect to metal ratios, were mixed and thoroughly milled in an agate mortar. The mixed powder was placed in an alumina crucible and calcined at 750°C for 15 hours in an $O_2$ environment. After this initial calcination the sample was pressed into a pellet and then heated up at 900°C for 2 days and 1100°C for two days in an $O_2$ environment with several intermediate grindings, and new pellets made after each grinding step.

3.2. Chemical composition
The chemical composition of the prepared ceramic BCTO samples was analyzed by energy-dispersive spectroscopy (EDS) using a JEOL 840A scanning electron microscope and INCA 4.07 (Oxford Instruments) software. The analyses performed on several particles showed that the concentration ratios of Ba, Co and Te are the stoichiometric ones within the instrumental resolution (0.05).

3.3. Second harmonic generation (SHG) measurements.
The materials were characterized by SHG measurements in reflection geometry, using a pulsed Nd:YAG laser ($\lambda$=1064nm). The SHG signal $I2\omega$ was measured from the polycrystalline samples relative to an $\alpha$-quartz standard at room temperature in the Q-switching mode with a repetition rate of 4 Hz.

3.4. Magnetic and dielectric measurements
The magnetization experiments were performed in a Quantum Design MPMSXL 5 T SQUID magnetometer. The magnetization (M) was recorded as a function of temperature (5-300 K) in 20 and 1000 Oe field using zero-field-cooled (ZFC) and field-cooled (FC) protocols.
Dielectric properties of BCTO ceramic samples were measured using ceramic disk (0.3mm thick) with silver electrodes fired on the both sides. The dielectric constant was derived from a R-5083 analyzer at a different frequencies.

**3**.5. Specific heat measurements
Specific heat measurements were performed using a relaxation method from T=2 K to 50 K on a Physical Properties Measurement System (PPMS6000) from Quantum Design Inc.

3.6. X-ray powder diffraction

The phase identification and purity of the powder sample was checked from X-ray powder diffraction (XRPD) patterns obtained with a D-5000 diffractometer using Cu $K_\alpha$ radiation. The ceramic sample of BCTO were crushed into powder in an agate mortar and suspended in ethanol. A Si substrate was covered with several drops of the resulting suspension, leaving randomly oriented crystallites after drying. The XRPD data for Rietveld analysis were collected at room temperature on a Bruker D8 Advance diffractometer (Ge monochromatized Cu $K_\alpha$ radiation, Bragg-Brentano geometry, DIFFRACT plus software) in the $2\theta$ range 10-152° with a step size of 0.02° (the counting time was 15 s per step). The slit system was selected to ensure that the X-ray beam was completely within the sample for all $2\theta$ angles.

3.7. Neutron powder diffraction

Because the neutron scattering lengths of Co and Te are different, the chemical composition of the B-site cations can be observed by neutron powder diffraction (NPD) with good precision ($b_{Co}$= 2.49 fm, $b_{Te}$ = 5.80 fm). The neutron scattering length of oxygen is comparable to those of the heavy atoms and NPD provide accurate information on its position and stoichiometry.

The neutron diffraction experiments on BCTO samples were performed at the Institute Laue-Langevin (Grenoble, France) on the powder diffractometer D1A (wavelength 1.91 Å) in the $2\theta$-range 10– 156,9° with a step size of 0.1°. The powdered sample was inserted in a cylindrical vanadium container. A helium cryostat was used to collect data in the temperature range 1.5-295 K. Nuclear and magnetic refinements were performed by the Rietveld method using the FULLPROF software [32]. The diffraction peaks were described by a pseudo-Voigt profile function, with a Lorentzian contribution to the Gaussian peak shape. A peak asymmetry correction was made for angles below 35° ($2\theta$). Background intensities were described by a polynomial with six coefficients. During the refinements the two octahedrally coordinated metal cations (Co and Te) were allowed to vary their occupation on the two possible metal sites. The magnetic structure was determined by magnetic symmetry analysis [32-34]. With the help of K-SEARCH software [32] the magnetic Bragg peaks can be indexed and the propagation vector can be found. The program BASIREPS [32], was used for calculation of the irreducible representations. Input information for this analysis was the nuclear space group, the propagation vector and the position of the independent magnetic ions. This program gave us to the command lines for the magnetic refinement in the frame of the FULLPROF software [32]

The magnetic structure was refined in space group P1 as an independent phase in which only $Co^{2+}$ cations were included. Several magnetic models were tried in the refinement, each employing one additional refinement parameter, corresponding to the magnitude of the magnetic moment. Each structural model was refined to convergence. The variant for which the structural refinement was stable and the R factors at minimum was chosen as the final model.

**4. Results**

4.1 Chemical composition

According to the elemental analyses performed on 20 different crystallites, the metal compositions of BCTO is $Ba_{0.99(2)}Co_{0.49(2)}Te_{0.52(2)}$, if the sum of the cations is assumed to be 2. These values are very close to the expected ratios and permit to conclude that the sample stoichiometry is the nominal one. The oxygen content, as determined by thermogravimetric analysis, is also in agreement with the $Ba_2CoTeO_6$ composition. The microstructure of the obtained powders, observed by scanning electron microscopy, reveals uniform and fine grain distribution.

4.2 Second harmonic generation (SHG) measurements

The SHG measurements at room temperature gave a negative result, thus testifying that at this temperature the BCTO compound probably possesses a centrosymmetric crystal structure. This sample still could be non-centrosymmetric, but at a level detectable only with sensitivities beyond $10^{-2}$ of quartz [35].

4.3 Magnetic measurements

The magnetic response is essentially linear with field at all measured temperatures, indicating paramagnetic/antiferromagnetic behaviour. Figure 1 shows the magnetic susceptibility vs. temperature at applied fields of 20 and 1000 Oe. In the inset the susceptibility measured in a magnetic field of 1000 Oe is plotted over an extended temperature range. The curve shows a pronounced maximum at about 18 K signaling a low temperature antiferromagnetic phase. The main frame shows the susceptibility in a field of 20 Oe at low temperatures. The ZFC and FC curves coalesce and the onset of antiferromagnetic order can from the temperature dependence of the quantity $d(\chi T)/dT$ (which mimics the temperature dependence of the magnetic specific heat) be estimated to occur at 15 K. This feature, and some associated complications, is further commented below in connection with the specific heat results. A Curie-Weiss analysis of the high-temperature susceptibility data (40K < T < 400 K) yields a surprisingly large value of $\theta \approx$ -180 K and an effective moment of p = 6.7 $\mu_B$. In the related $Ca_2CoTeO_6$ ($T_N$=10K) and $Sr_2CoTeO_6$ ($T_N$=15K) compounds, p = 5.5 $\mu_B$ is obtained [36]. The present p value thus differs significantly from those of $Ca_2CoTeO_6$ and $Sr_2CoTeO_6$. It is close to the value theoretically expected for free $Co^{2+}$ in a high-spin $3d^7$ configuration, (S=3/2, L=3, p = 6.54 $\mu_B$), suggesting no orbital quenching. Interestingly more differences with the $Ca_2CoTeO_6$ and $Sr_2CoTeO_6$ compounds can be observed from the Curie-Weiss analysis. The frustration parameter $f = -\theta/T_N$ gives a measure of the frustration of the antiferromagnetic systems. f amounts to 7.8 for $Ca_2CoTeO_6$ [36], and increases to 9.7 for $Sr_2CoTeO_6$ [36]. In BCTO, f is even larger and amounts to 12, indicating an even greater influence of competing magnetic interaction and increased magnetic frustration.

4.4 Dielectric measurements

The temperature dependence of the dielectric constant $\varepsilon$ was measured in the temperature range 100-450K (see Figure 2). A broad and frequency dependent maximum on the $\varepsilon$ vs. T curve was observed near 280 K. The position of the dielectric maximum shifts to higher temperatures and decreases in amplitude with increasing frequency, thus indicating that the possible phase transition is diffuse and that the relaxation component makes a remarkable contribution to the spontaneous polarization if it exists. An additional investigation did not show a dielectric hysteresis loop up to breakdown field values, which indicates a possible antiferroelectric nature of this anomaly around 280 K. Detailed evaluations of the dielectric characteristics of BCTO dense ceramics in broad temperature and frequency ranges are in progress.

4.5 Specific heat

The temperature dependence of the heat capacity is shown in Figure 3 (plotted as C/T). Below a temperature of 18 K, the heat capacity start to increase with decreasing temperature until it reaches a maximum near about 12 K. This suggests a large magnetic entropy change around these temperatures, associated with a second order magnetic phase transition in agreement with the magnetization data. The broadness of the peak-like feature in the heat capacity might reflects the above mentioned magnetic frustration. Considering the baseline defined by the high-temperature C/T data indicated in the inset of Figure 3, we can estimate that the peak height corresponds to at least $\Delta C = 6.3$ J/K/mol, i.e. $\Delta C = 0.76$ R. By integrating the C/T data after subtraction of the baseline, a sizable minimum value of the entropy change associated with the magnetic transition $\Delta S = 0.18$ R can be estimated.

4.6 X-ray powder diffraction

The first crystallographic characterization of BCTO compound was performed by XRPD analysis at room temperature which showed that the prepared samples formed powders with hexagonally-distorted perovskite structure (see Figure 4). XRD data for BCTO and $Ba_2NiOsO_6$ are very similar suggesting that the compounds belong to the same structural type. An attempt to index the diffractograms with a $=\sqrt{2}a_p$ and c $=2\sqrt{3}a_p$ ($a_p$-parameter of primitive cubic perovskite unit) gave satisfactory result and a unit cell with lattice parameters $a =5.8020(3)$ Å, $c =14.2705(8)$Å was found, no impurity phases were detected. The systematic absences observed in the XRD pattern are compatible with several space groups. Attempts to fit a structural model were performed only for centrosymmetric space groups. The centrosymmetric space group P-3m gave the best fit with the smallest number of refined parameters. The starting model for the crystal structure of BCTO was the structure of hexagonal $Ba_2NiOsO_6$ as determined in [37]. Refinements of the site occupancy factors of the oxygen atoms did not reveal significant changes from full occupancy (within three standard deviations) and was therefore fixed at unity. The final Fourier difference synthesis calculation showed no significant maxima. The lattice parameters for BCTO are in reasonable agreement with those reported in [29]. There is, however, some disagreement with the cell parameters of the ceramic samples of BCTO reported in [28]. The difference in lattice parameters could be explained by the different sample preparation methods. No additional deviation from the trigonal symmetry could be observed by XRPD.

4.7 Neutron powder diffraction

We started to refine the crystal structure of BCTO sample using NPD data at 300 K. The magnetic contribution could be evaluated from the comparison of the data measured above and below $T_N$ (see Figure 5).

To confirm the space group found from the refinements of the X-ray data several centrosymmetric trigonal space groups were initially considered. Different models based on partial ordering between Co and Te cations in the B- sublattice (*1a* and *2d* sites) were also tested. Rietveld refinements were carried out in all space groups, but a clearly superior fit was obtained using space group P-3m with the structural model previously proposed from XRPD data. The obtained atom positions are very similar, but we were able to determine more accurately the oxygen positions due to the characteristics of the neutron scattering. A model describes an ordered distribution of Co and Te in the B-site. No vacancies were observed in the cationic or in the anionic substructures. Accordingly, the chemical composition seems to be very close to the nominal one and therefore, the cobalt oxidation state can be assumed to be two.

The atomic coordinates and other relevant parameters are tabulated in Table 1. Selected bond lengths and angles are listed in Table 2. Figure 6 shows a view of the crystal structure of BCTO. The structure consists of dimers of face-sharing octahedra separated by single corner sharing octahedra. It is clear that there are two distinct transition metal sites; B(1) shares only vertices with adjacent octahedra whilst B(2) shares both vertices and faces with neighbouring octahedra.

The Co(1)$O_6$ and Te(1)$O_6$ octahedra are constrained by the structural symmetry to have six equal bond lengths and hence they are highly regular in shape, while the Co(2) and Te(2) octahedra are less regular. There are three shorter distances and three longer ones. Adjacent Co(2)/Te(2) octahedra are sharing one face in a dimer-like arrangement with Co-Te separation across the shared face of 2.7Å. These dimer-like units connect to the Co(1) and Te(1) octahedra by corner sharing, showing Co-O-Te bond angles close to 178º. The barium cations are located in 12-fold coordination sites. These coordination environments are typical for Ba in close-packed oxygen arrays.

NPD measurements were performed at different temperatures between 5 and 295 K in order to check whether any phase transition occurred or in BCTO, and in order to follow the evolution of the nuclear and magnetic structures at low temperatures. The diffraction profiles are shown in Figure 7. There is no evidence of any major change in the crystal structure between 5 K and 295 K.

A comparison of the NPD powder patterns shows that some magnetic scattering exists below 20 K. Using the Rietveld method the nuclear and magnetic structures were determined.

As discussed in relation to the magnetic properties, the magnetic susceptibility measurements indicate an initial antiferromagnetic transition at about 15 K. The NPD patterns of BCTO obtained below the Neel temperature show the presence of additional peaks of magnetic origin at low-angles. As the temperature decreases, the intensity of the magnetic peaks increases. For the refinement of the high resolution data with $T < 20$ K, the presence of magnetic peaks has to be taken into account. The magnetic Bragg peaks were indexed using a nuclear unit cell doubled in the *a*-direction; the magnetic propagation vector was proposed to be $k = [1/2, 0, 0]$ and we can conclude that the magnetic structure is antiferromagnetic and commensurate with the nuclear lattice. The presence of two possible magnetic sites for Co cations in the nuclear unit cell prompted us to use magnetic symmetry analysis for the determination of the allowed irreducible representations of $k = [1/2, 0, 0]$ in space group *P*-3m. The different magnetic arrangements were then investigated, checking all the proposed basis functions of the irreducible representations of the $P$-3m space group for the propagation vector $k$ = [1/2, 0, 0]. For each combination of basis function, we have carried out a systematic comparison between the observed and calculated neutron diffraction patterns at 5 K.

After checking all of the possible magnetic modes obtained by BASIREPS program (see Table 3), the best agreement with the experimental data was obtained for the collinear antiferromagnetic ordering with the largest component of the magnetic moments lying along $c$ direction given by the basis vectors of the irreducible representation $\Gamma_3$. A view of the magnetic structure associated to $\Gamma_3$ is presented in Figure 8. The modulus of the magnetic moment of the Co cations is 2.34(5) $\mu_B$ and the canting angle ($\varphi$) of the magnetic moment from the c-axis is around 24.5°. The Co spins of the layers perpendicular to the c-axis are antiferromagnetically coupled in each layer.

The analysis of the obtained magnetic structure shows that several magnetic interactions have to be considered in this system. The first neighbour interaction seems to be a direct exchange between $Co^{2+}$ cations, but as the shorter Co-Co distances are 5.2 Å (Co(2)-Co(2)) and 5.7 Å (Co(1)-Co(1)), the direct overlap of the unpaired electron wave functions should be rather weak. The second kind of magnetic interactions should be of supersuperexchange type where two neighboring Co atoms interact through two oxygen atoms (or through Te and oxygen atoms). Such type of interaction takes place for the next-nearest neighbors with distances near 6.1 Å. The magnetic structure is the result of the competing magnetic interactions, and it presents antiferromagnetic alignment of the magnetic moments of the two cobalt ions. Concerning the spin state of Co ions, the refinement of the magnetic structure for BCTO revealed a reduction in the observed magnetic moment (2.34(5) $\mu_B$) in comparison with the expected value of 3.0 $\mu_B$. The reduction of the ordered magnetic moment compared with the pure ionic configuration can be due to a combination of the covalence effects.

## 4. Discussion

The refined atomic coordinates and bond distances for BCTO (see Tables 1 and 2) confirm the basic structural features of the proposed 6L-$BaTiO_3$ structure (Figure 6). It comprises close-packed $AO_3$ layers with the B-type cations occupying two different types of octahedral sites: Co(1) and Te(1) occupies perovskite-like corner-shared octahedra while Co(2) and Te(2) occupies pairs of face-shared octahedra, $B_2O_9$. There are three different oxygen sites which occupied cubic and hexagonal-stacked $BaO_3$ layers and separated the pairs of face-shared octahedra. It may be noted that the O(1)-O(1) distances within the shared edge are comparatively short (~ 2.60Å), thus helping to "shield" the Co-Te repulsion. The average Co(1)-O bond length is very similar to that found in other oxides containing $Co^{2+}$ [39, 40] and close to the $Co^{2+}$-O length (Å) calculated from Shannon´s ionic radii [41] for a 6 coordinated high spin state. The estimated Co(2)-O distances differ from the expected ones. The possible Co-O-Te superexchange pathways are almost linear (the Co-O-Te angle is about 178°). In order to get some insight into the cation distribution, we carried out bond-valence sum calculations according to Brown´s model [42, 43] which gives a relationship between the formal valence of a bond and the corresponding bond lengths. In non-distorted structures, the bond valence sum rule states that the valence of the cation ($V_i$) is equal to the sum of the bond valences ($v_{ij}$) around this cation. From individual cation-anion distances the valences of cations were calculated. The Co(1) and Co(2) cations exhibit the valences +2.14(1) and +1.95(1), respectively (see Table 4), which slightly differ from the expected +2 valence of cobalt in this compound. The valence of the Te cations is somewhat lower than +6 (+5.93(1) for Te(1) and 5.97(1) for Te(2)).

Two major factors will determine the arrangement of different cations at the B-sites in the perovskite structure. The ions can be either ordered or statistically distributed (disordered), and the controlling factors are the difference in ionic charge and in ionic size [15]. It has been shown [3, 15-17] that the magnetic properties of complex perovskites strongly depend on the B-site ordering. The ionic radii of $Co^{2+}$ and $Te^{6+}$ are not so similar and therefore we may say that this combination of cations in general favor an ordered arrangement.

It can be noted that other double perovskites with an ordered cubic structure from the barium series of compounds, namely $Ba_2CoMO_6$ (M=W, Mo, Re, U) [44-50], also order antiferromagnetically at low temperatures (see Table 5), all with somewhat different but still low values of $T_N$. It is important to understand the possible influence of the M cations on this difference. The difference in the size of different $M^{6+}$ cations are not large enough to explain the different values of $T_N$, but as a tendency, the value of $T_N$ decreases with increasing $M^{+6}$ cation size. It should be particularly emphasized that this analysis involves difficulties from the fact that available structural and magnetic data are very restricted. Consequently, it is not possible to find a simple explanation for the different values of $T_N$ based on only crystallochemical considerations using the tolerance factor *t*. If we consider other series of materials, namely $A_2CoTeO_6$ (A= Ba, Sr, Ca) [36] (see Table 6) it is clear that in the case of monoclinically distorted phases (for Sr and Ca) $T_N$ slightly increases with an increase of the $A^{2+}$ cation size, but the same difficulties arise due to the limited crystallographic information presently available for A=Pb and Cd. Detailed additional investigations of complex metal oxides with different B-site cations would be necessary in order to clarify the possible correlations between structure and magnetic properties. In analogy with other Co-based perovskites [15, 40, 44, 45], however, we may assume that $T_N$ is strongly related with the value of the Co-Co distance, the Co-O-Co angle and with the type of lattice distortions (see Tables 5, 6). At the same time it is clear that the degree of crystallographic disorder at Co-sites (so called anti-site disorder) and /or oxygen deficiency should vary with the type of M cations. For instance, in our previous research [51] it was shown that a small oxygen deficiency had a pronounced effect on the structural parameters and magnetic properties of complex metal oxides adopting a hexagonal perovskite structure. In addition, the effects of frustration due to competing ferro- and antiferromagnetic interaction severely lower the antiferromagnetic transition temperature compared to the interaction strength indicated by the measured Weiss temperature.

All the members of the series $A_2CoTeO_6$ (A=Ba,Pb,Sr,Ca,Cd) also belong to the family of double perovskites with a different structural distortion (tolerance factor) and ordering of B-type cations. It has earlier been mentioned in [15,29] that the temperature of (anti)ferroelectric phase transition $T_C$ for $A_2CoTeO_6$ perovskites is increasing with decreasing size of the A-type cations (see Figure 9). As the effective size of A-type cation decreases from Ba to Cd, the size of the A cation is too small for the 12-fold site within a

BO$_6$ octahedral framework. BO$_6$ octahedra tilt to optimize the distorted A-O distances which eventually results in a decrease in its coordination number. In the case A=Pb,Sr,Ca,Cd this tilting does not disrupt the corner-sharing connectivity present in the ideal cubic perovskite. But in the case of BCTO a tolerance factor larger than unity implies that Ba cations are too large to be accommodated in the A-site and hexagonal perovskite lattice which gives more space to large Ba cations is formed. But, on the other hand, some octahedral share faces leading to short distance between B-type cations inside these face-sharing octahedra. While the distance between the B-type cations in two corner-sharing polyhedra is about 4 Å, the B-B distance over the common face is as short as 2.7 Å, hence comparable to the B-B distance in metallic compounds. Face sharing of octahedra is unexpected in purely ionic oxides and this covalency of the B-O bonds is important for stabilizing the hexagonal structure. Covalency implies directional bonding that will reduce the Coulomb repulsion between adjacent metal atoms.

It is well-known [15,30,31] that there are only three structural degrees of freedom, which may be responsible for lattice distortion in complex perovskites: a) cation shifts, b) distortions of the oxygen polyhedra coordinating the cations in the A and B-sublattices and c) tilting of the oxygen octahedra. If one takes into account the values of the polyhedra distortions (see Table 4), the ferroic properties of BCTO may be connected with the Co(2) and Ba(3)-sublattices, which have significant values of the off-center cation displacements from their polyhedral centers. At room temperature it was found, using a method proposed in [52], that the volume of the Ba(3) polyhedra (59.7 Å$^3$) is larger than the volume for the Ba(1) and the Ba(2) polyhedra (57.8 Å$^3$) and (57.3 Å$^3$) respectively. In the Co-sublattice, Co(1) cations have a certain degree of "freedom" in comparison with the Co(2) ions (13.6 and 12.3 Å$^3$, respectively). It is worth to notice that the volume of the Te(1) and Te(2) polyhedra are practically identical (9.5(1) and 9.4(1) Å$^3$, respectively). At the same time the off-center displacements of cations are zero for Co(1) and Te(1) following the symmetry restrictions, while these values for the Ba(3) and Co(2) cations are significantly greater (0.118 and 0.126 Å, respectively).

The Co(1) and Te(1) octahedra are constrained by the structural symmetry to have six equal bond lengths with the oxygens. The Co(2) and Te(2) octahedra are less regular. There are three shorter distances and three longer ones due to a shift of the cations within its octahedral blocs to reduce repulsion between them.

The position in the corner-shared octahedra is ordinary occupied by the cations with low charge state (+2 or +3), whereas the face-shared octahedra contain high charge cations (+5 or +6) [15, 17]. The degree of cation ordering is always very high, which means that the cation distribution meets the requirements of the structure stability. In the case of BCTO a fully ordered model was realized when the two B-type cations were distributed between the corner-shared octahedra and the face-shared octahedra.

We are planning to reinforce the magnetic interactions in BCTO by introducing of Co excess in Te-sublattice (thus creating some anti-site disorder). We expect such substitution to promote nearest neighbor Co-O-Co antiferromagnetic spin coupling and thus raise the value of $T_N$. Both a change of the stoichiometric ratio between Co and Te cations and lowering of the temperature for solid-state sintering procedure can help in this direction. This research is in a progress.

## 5. Concluding remarks

The fully ordered 6L-perovskite $Ba_2CoTeO_6$ was synthesised and structurally characterised applying the Rietveld analysis of XRPD and NPD data at different temperatures. Both Co and Te cations occupy the face-sharing octahedra (*2d*) and the corner-sharing octahedra (*1a*) sites. The precise metal-oxygen distances derived from NPD data made it possible to calculate the valences of the cations and the distortion of their polyhedra. A model for the antiferromagnetic structure with the Co moments directed along the c-axis was proposed. Structural and magnetic features of BCTO are considered and compared with those of other quaternary complex oxides. Further detailed dielectric measurements are needed before the possible magnetoelectric properties of BCTO can be fully understood. Moreover, the role of the anion sublattice, which is crucial for the understanding of magnetoelectric phenomena, will be investigated in more detail.

## Acknowledgements

Financial support of this research from the Royal Swedish Academy of Sciences, the Swedish Research Council (VR) and the Russian Foundation for Basic Research is gratefully acknowledged.

Table 1 Summary of the results of the structural refinements of the Ba$_2$CoTeO$_6$ sample using XRD and NPD data.

| Experiment | XRD | NPD | NPD |
|---|---|---|---|
| T,K | 295 | 5 | 295 |
| a (Å) | 5.8020(3) | 5.7884(2) | 5.7996(2) |
| c (Å) | 14.2705(7) | 14.2497(5) | 14.2658(5) |
| s.g. | P-3m | P-3m | P-3m |
| Ba(1) 2c(0,0,z) | | | |
| z | 0.2511(8) | 0.2509(7) | 0.2483(8) |
| B[Å]$^2$ | 0.91(3) | 0.68(2) | 0.87(2) |
| Occupancy | 1.02(3) | 0.996(5) | 0.996(5) |
| Ba(2) 2d(1/3,2/3,z) | | | |
| z | 0.0956(8) | 0.0844(7) | 0.0864(8) |
| B[Å]$^2$ | 0.75(3) | 0.52(2) | 0.65(2) |
| Occupancy | 0.995(6) | 1.010(8) | 1.009(6) |
| Ba(3) 2d(1/3,2/3,z) | | | |
| z | 0.4001(8) | 0.3946(7) | 0.3986(8) |
| B[Å]$^2$ | 0.82(3) | 0.61(2) | 0.74(2) |
| Occupancy | 1.007(8) | 0.992(9) | 0.994(7) |
| Co(1) 1a(0,0,0) | | | |
| B[Å]$^2$ | 0.25(4) | 0.16(3) | 0.26(3) |
| Occupancy | 0.994(9) | 1.005(6) | 1.007(6) |
| Co(2) 2d(1/3,2/3,z) | | | |
| z | 0.6573(8) | 0.6560(7) | 0.6400(7) |
| B[Å]$^2$ | 0.36(4) | 0.26(3) | 0.35(3) |
| Occupancy | 1.006(9) | 0.995(8) | 1.008(6) |
| Te(1) 1b(0,0,1/2) | | | |
| B[Å]$^2$ | 0.34(4) | 0.12(3) | 0.19(3) |
| Occupancy | 1.010(9) | 0.991(9) | 0.993(8) |
| Te(2) 2d(1/3,2/3,z) | | | |
| z | 0.8375(8) | 0.8370(7) | 0.8351(7) |
| B[Å]$^2$ | 0.19(2) | 0.19(3) | 0.28(3) |
| Occupancy | 0.990(9) | 1.009(9) | 1.003(6) |
| O(1) 6i(x,-x,z) | | | |
| x | 0.5138(7) | 0.5176(6) | 0.5121(6) |
| z | 0.2501(9) | 0.2469(7) | 0.2428(7) |
| B[Å]$^2$ | 0.86(4) | 0.72(3) | 0.86(3) |
| Occupancy | 1.02(4) | 0.994(9) | 0.994(7) |
| O(2) 6i(x,-x,z) | | | |
| x | 0.8229(7) | 0.8239(6) | 0.8236(6) |
| z | 0.0834(8) | 0.0875(7) | 0.0872(7) |
| B[Å]$^2$ | 0.93(5) | 0.69(3) | 0.85(3) |
| Occupancy | 0.98(3) | 1.012(9) | 1.013(8) |

| O(3) 6i(x,-x,z) | | | |
|---|---|---|---|
| x | 0.8380(7) | 0.8395(7) | 0.8423(7) |
| z | 0.4196(7) | 0.4243(5) | 0.4239(5) |
| $B[Å]^2$ | 0.91(5) | 0.75(1) | 1.02(3) |
| Occupancy | 1.04(4) | 1.004(9) | 1.011(8) |
| $R_p$,% | 5.42 | 4.87 | 5.26 |
| $R_{wp}$,% | 6.84 | 6.12 | 6.57 |
| $R_B$,% | 5.41 | 4.36 | 5.29 |
| $R_{mag}$,% | - | 8.62 | - |
| $\mu(\mu_B)$Fe | - | 2.34(5) | - |
| $\chi^2$ | 2.19 | 2.24 | 2.31 |

Table 2 Selected bond distances from neutron powder refinements of the $Ba_2CoTeO_6$ sample at various temperatures

| Bonds[Å] | 5K | 295K |
|---|---|---|
| Ba(1)　O(1) x6 | 2.900(3) | 2.904(4) |
| 　　　　O(2) x3 | 2.923(4) | 2.902(5) |
| 　　　　O(3)x3 | 2.948(6) | 2.964(7) |
| Ba(2)　O(1)x3 | 2.963(3) | 2.7864(4) |
| 　　　　O(2)x6 | 2.896(4) | 2.902(5) |
| 　　　　O(2) x3 | 2.912(6) | 2.936(7) |
| Ba(3)　O(1)x3 | 2.801(4) | 2.857(4) |
| 　　　　O(3)x6 | 2.926(6) | 2.924(6) |
| 　　　　O(3)x3 | 3.108(6) | 3.086(6) |
| Co(1)　O(2)x6 | 2.161(5) | 2.165(6) |
| Co(2)　O(1)x3 | 2.037(5) | 2.281(6) |
| 　　　　O(3)x3 | 2.076(5) | 1.986(6) |
| Te(1)　O(3)x6 | 1.937(4) | 1.921(5) |
| Te(2)　O(1)x3 | 1.914(4) | 1.910(6) |
| 　　　　O(2)x3 | 1.908(5) | 1.928(6) |
| Co(1)-O(1)-Te(2)° | 178.8(7) | 178.1(8) |
| Co(2)-O(3)-Te(1)° | 179.6(7) | 177.3(8) |
| Co(2)-O(1)-Te(2)° | 81.7(7) | 79.8(8) |

Table 3 Irreducible representations of the small group $G_K$ obtained from the space group P-3m for k=(0.5 0 0 ) and the corresponding basis vectors. The symmetry elements are written according to [37].

| Ired. repres. | Basis vectors | $h_1$ | $h_7$ | $h_{13}$ | $h_{19}$ | Cations | | |
|---|---|---|---|---|---|---|---|---|
| | | | | | | $Co_1(0,0,0)$ | $Co_{21}(1/3,2/3,0.656)$ | $Co_{22}(2/3,1/3,0.344)$ |
| $\Gamma_1$ | $V_{11}$ | 1 | 1 | 1 | 1 | [ 0 -1 0] | [ 0 -1 0] | [ 0 1 0] |
| $\Gamma_2$ | $V_{21}$ | 1 | 1 | -1 | -1 | - | [ 2 1 0] | [ 2 1 0] |
| | $V_{22}$ | | | | | | [ 0 0 1] | [ 0 0 1] |
| $\Gamma_3$ | $V_{31}$ | 1 | -1 | 1 | -1 | [ 2 1 0] | [ 2 1 0] | [ -2 -1 0] |
| | $V_{32}$ | | | | | [ 0 0 1] | [ 0 0 1] | [ 0 0 -1] |
| $\Gamma_4$ | $V_{41}$ | 1 | -1 | -1 | 1 | - | [ 0 -1 0] | [ 0 -1 0] |

Table 4 Polyhedral analysis of the $Ba_2CoTeO_6$ sample at 295K (c.n. - coordination number, x – shift from centroid, ξ- average bond distance, V- polyhedral volume, ω- polyhedral volume distortion.

| Cation | cn | x[Å] | ξ [Å] | V[Å³] | ω | Valence |
|---|---|---|---|---|---|---|
| Ba(1) | 12 | 0.027 | 2.919 | 57.8(1) | 0.084 | 2.20 |
| Ba(2) | 12 | 0.012 | 2.900 | 57.2(1) | 0.075 | 2.28 |
| Ba(3) | 12 | 0.118 | 2.940 | 59.6(1) | 0.076 | 2.07 |
| Co(1) | 6 | 0 | 2.166 | 13.6(1) | 0 | 2.14 |
| Co(2) | 6 | 0.126 | 2.110 | 12.3(1) | 0.016 | 1.95 |
| Te(1) | 6 | 0 | 1.921 | 9.5(1) | 0 | 5.93 |
| Te(2) | 6 | 0.017 | 1.919 | 9.4(1) | 0.08 | 5.97 |

Table 5 Crystallographic and magnetic data for $Ba_2CoMO_6$ perovskites (c.n. - coordination number, AS-anti-site disorder)

| $M^{6+}$ | Mo[45] | W[45] | W[46] | Te | Re[47] | Re[48] | Re[49] | U[50] |
|---|---|---|---|---|---|---|---|---|
| r[Å] | 0.59 | 0.60 | 0.60 | 0.56 | 0.55 | 0.55 | 0.55 | 0.73 |
| a[Å] | 8.086(1) | 8.108(1) | 8.103(1) | 5.799(1) | 8.072(1) | 8.086 | 8.051 | 8.374(1) |
| c[Å] | | | | 14.266(1) | | | | |
| s.g. | Fm3m | Fm3m | Fm3m | P-3m | Fm3m | Fm3m | Fm3m | Fm3m |
| c.n. | 6 | 6 | 6 | 6 | 6 | 6 | 6 | 6 |
| AS | No | No | Yes | No | Yes | No | No | No |
| $T_N$ | 27 | 19 | 17 | 15 | 25 | 40 | 41 | 9 |
| property | AFM | AFM | AFM | AFM | AFM | AFM | AFM | FM |
| Co-O[Å] | 2.114(1) | 2.129(1) | - | 2.037(5)-2.161(5) | 2.102(1) | - | 2.093(1) | 2.067(1) |

Table 6 Crystallographic and magnetic data for $A_2CoTeO_6$ perovskites
(AS-anti-site disorder)

| $M^{+6}$ | Ca[36] | Sr[36] | Ba |
|---|---|---|---|
| r[Å] | 1.34 | 1.44 | 1.61 |
| a[Å] | 5.4569(1) | 5.6417(1) | 8.103(1) |
| b[Å] | 5.5904 | 5.6063(1) | 5.799(1) |
| c[Å] | 7.7399 | 7.9239(1) | 14.266(1) |
| β(deg) | 90.24 | 90.12 | - |
| V[Å$^3$] | 236.1 | 250.6 | 415.5 |
| s.g. | P2$_1$/n | P2$_1$/n | P-3m |
| AS | No | No | No |
| T$_N$ | 10 | 15 | 15 |
| property | AFM | AFM | AFM |
| <A-O>[Å] | 2.794(3) | 2.815(10) | 2.918(6) |
| <Co-O>[Å] | 2.110(2) | 2.067(7) | 2.149(5) |
| <Te-O>[Å] | 1.930(2) | 1.937(7) | 1.921(6) |
| Co-O-Te (deg) | 148.1-151.5 | 163.5-169.3 | 177.3-178.1 |

**Figure captions**

Figure 1 Temperature dependence of the susceptibility of $Ba_2CoTeO_6$. The main frame shows low temperature results in a field of 20 Oe and the inset a wider temperature range in 1000 Oe.

Figure 2 Temperature dependence of dielectric permittivity of $Ba_2CoTeO_6$

Figure 3 The specific heat ($C_p/T$) vs. temperature of $Ba_2CoTeO_6$. The baseline used to determine the entropy change at the transition is indicated by the continuous line in the inset.

Figure 4 The observed, calculated, and difference plots for the fit to the XRPD pattern of $Ba_2CoTeO_6$ after Rietveld refinement of the nuclear structure at 295K.

Figure 5 The observed, calculated, and difference plots for the fit to the NPD patterns of $Ba_2CoTeO_6$ after Rietveld refinement of the nuclear and magnetic structure at different temperatures: 295 K (a) and 10 K (b).

Figure 6 The 6L-perovskite structure of $Ba_3CoTeO_6$.

Figure 7 Temperature evolution of NPD patterns of $Ba_2CoTeO_6$ (the magnetic reflections are indicated by *).

Figure 8 The magnetic structure of $Ba_2CoTeO_6$. Diamagnetic ions are omitted.

Figure 9 Variation of the ferroic phase transition temperatures ($T_C$) for perovskite compounds of $A_2Co^{2+}TeO_6$ ($A^{2+}$=Ba, Sr, Pb, Ca,Cd) versus the $A^{2+}$-cation radius..

Author's addresses

***Sergey A. Ivanov***

Department of Inorganic Materials, Karpov' Institute of Physical Chemistry, Vorontsovo pole,10 105064 Moscow K-64, Russia
**E-mail**: ivan@cc.nifhi.ac.ru

***Per Nordblad, Roland Mathieu,***

Department of Engineering Sciences, The Ångstrom Laboratory , Uppsala University, Box 534, 751 21 Uppsala, Sweden
**E-mail**: per.nordblad@angstrom.uu.se

roland.mathieu@angstrom.uu.se

***Roland Tellgren***

Department of Inorganic Chemistry, The Ångstrom Laboratory, Box 538, University of Uppsala, SE-751 21, Uppsala , Sweden
**E-mail**: roland.tellgren@mkem.uu.se

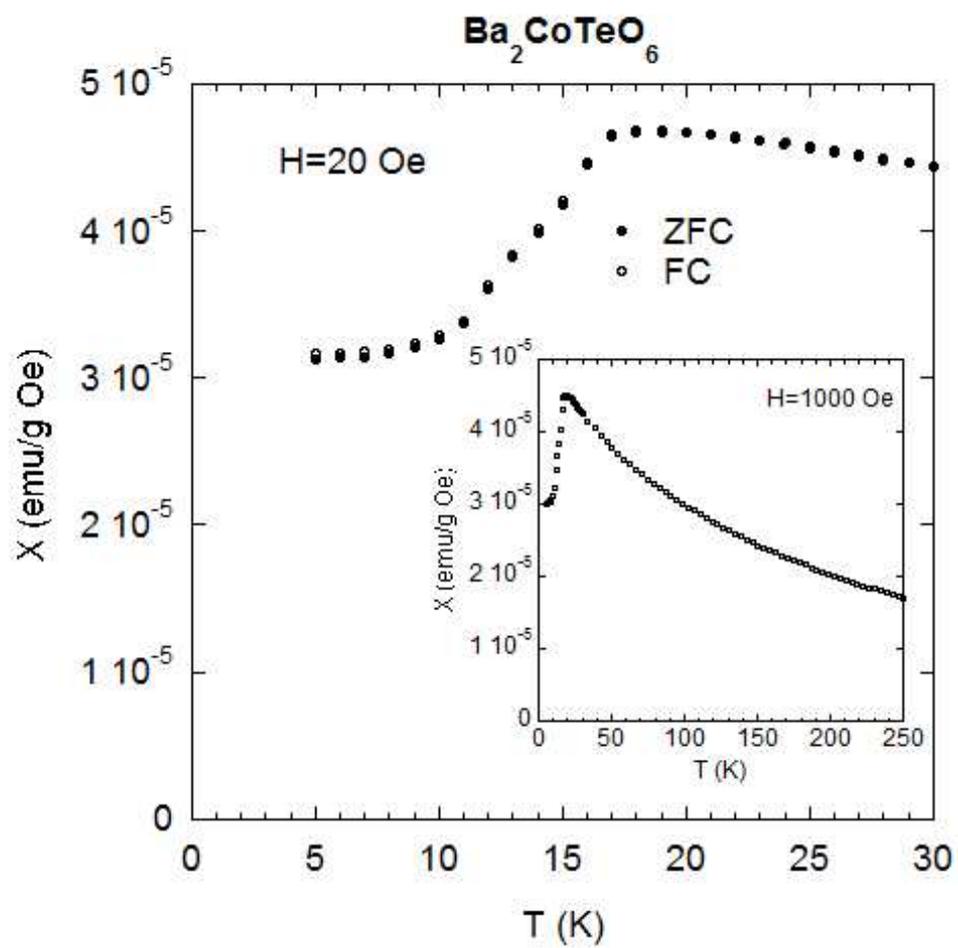

Figure 1

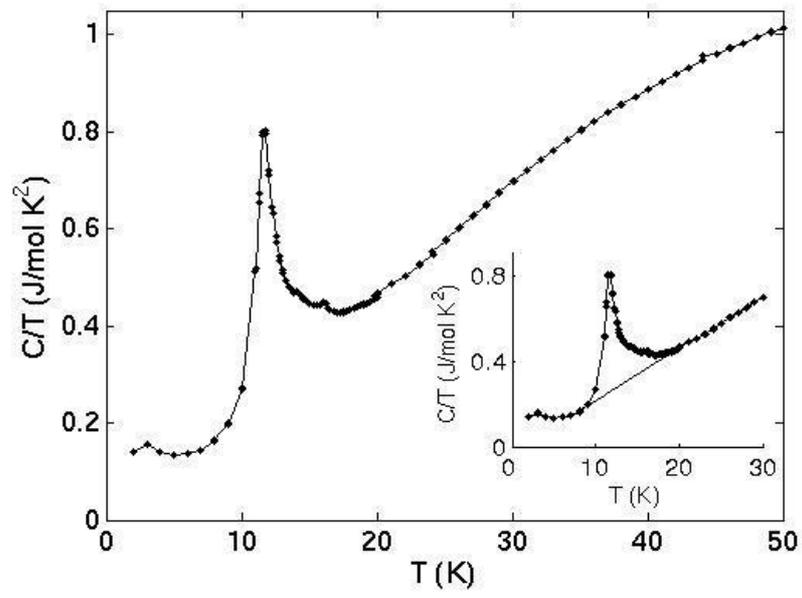

Figure 2

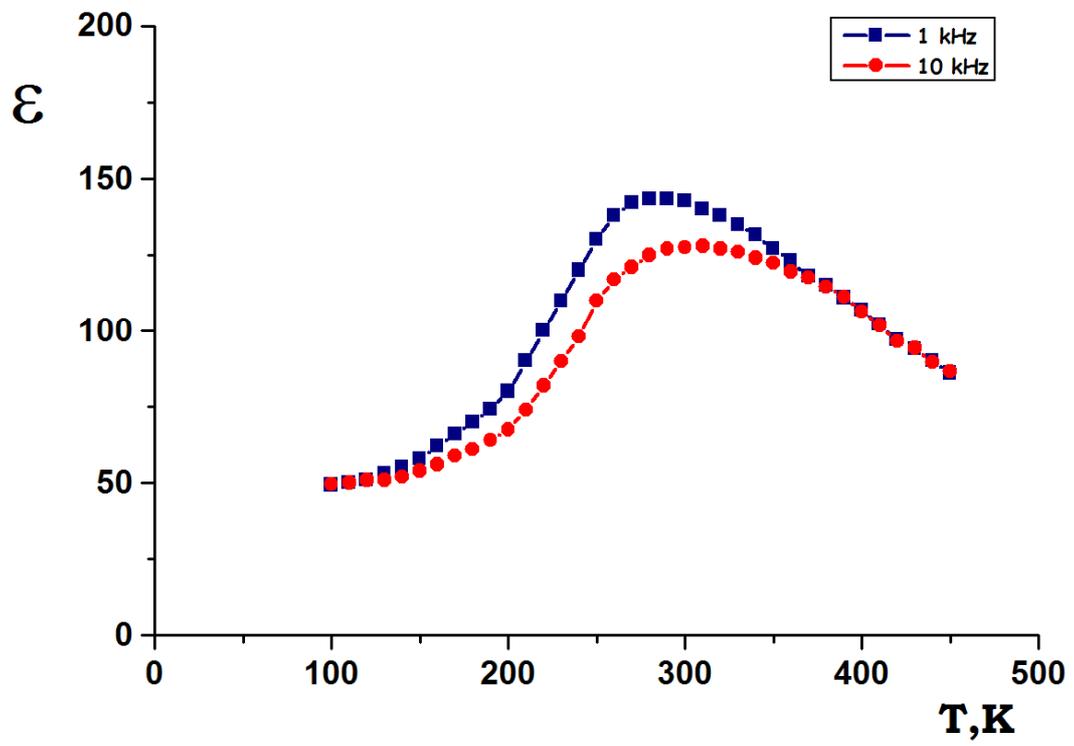

Figure 3

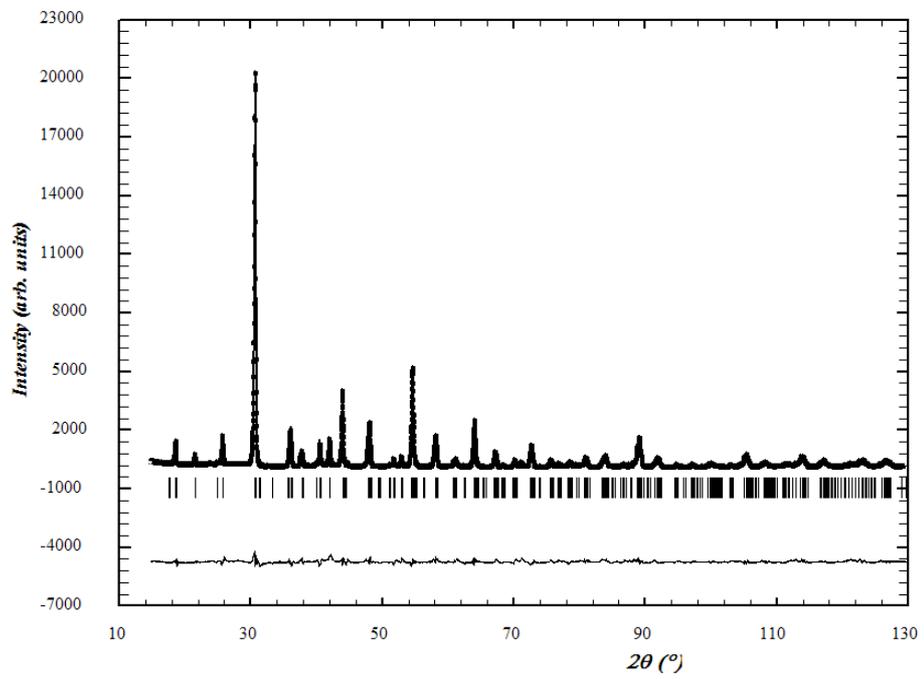

Figure 4

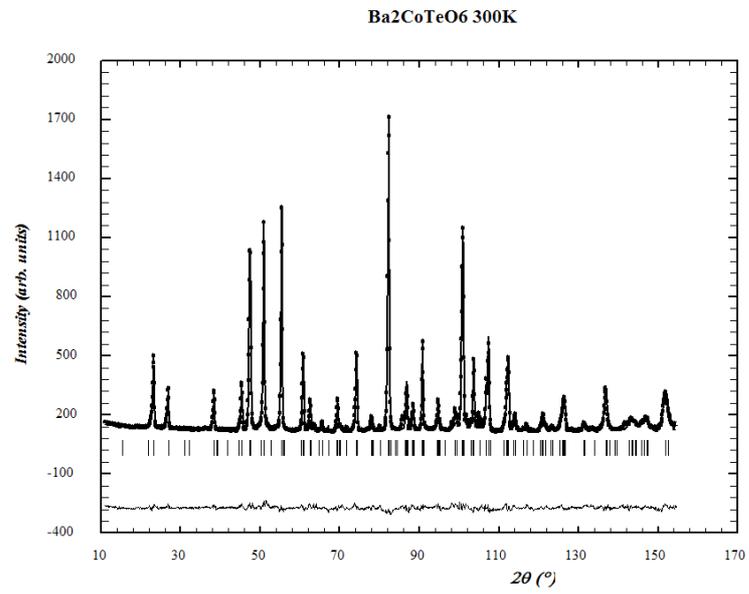

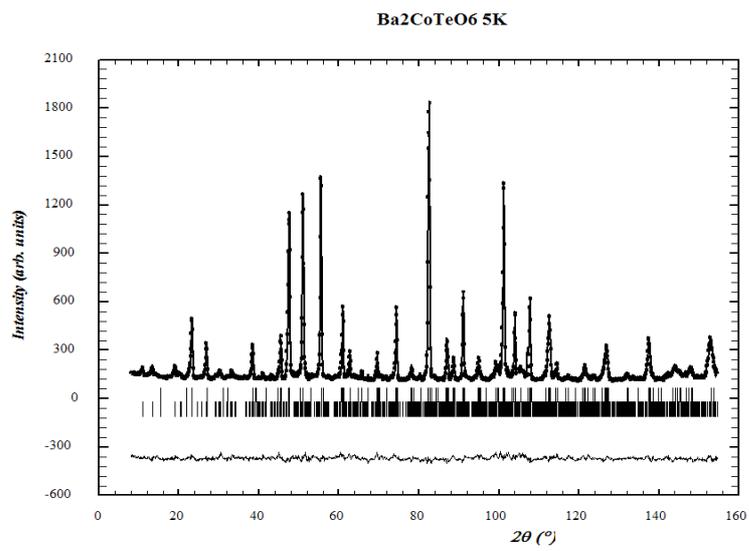

Figure 5

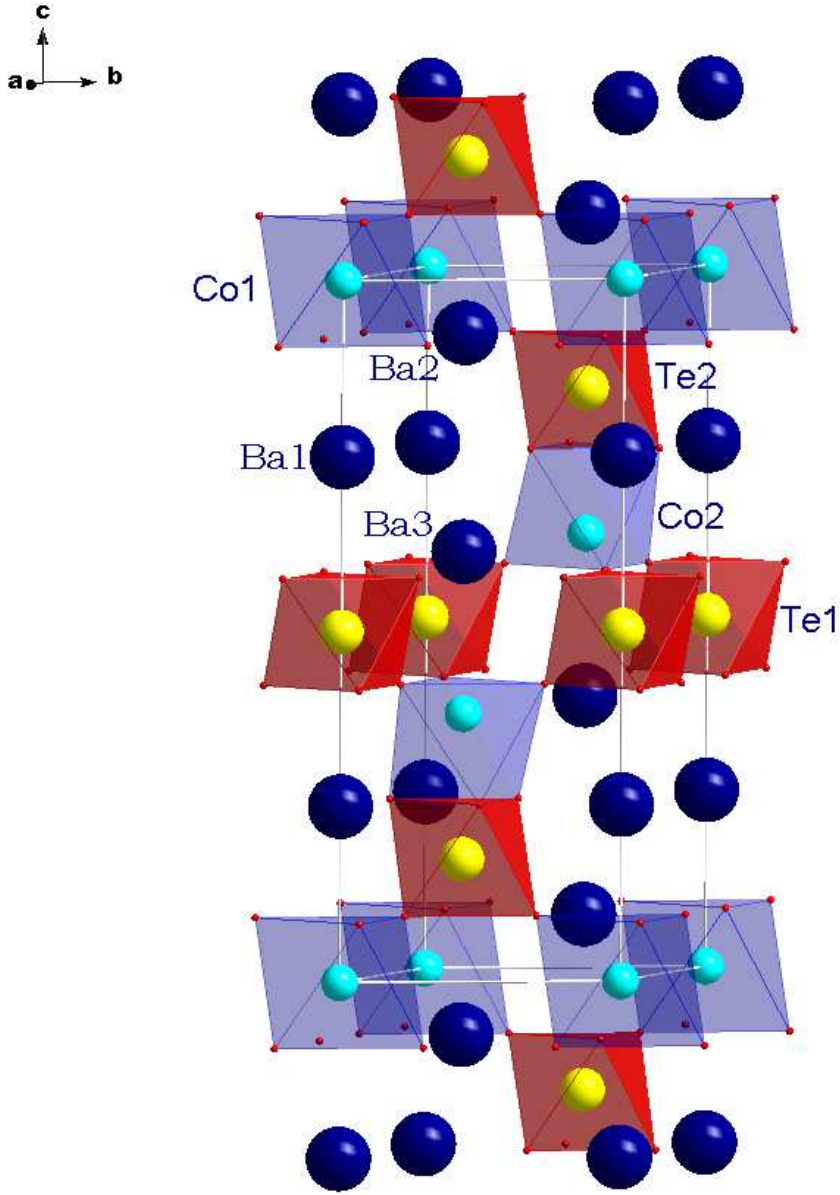

Figure 6

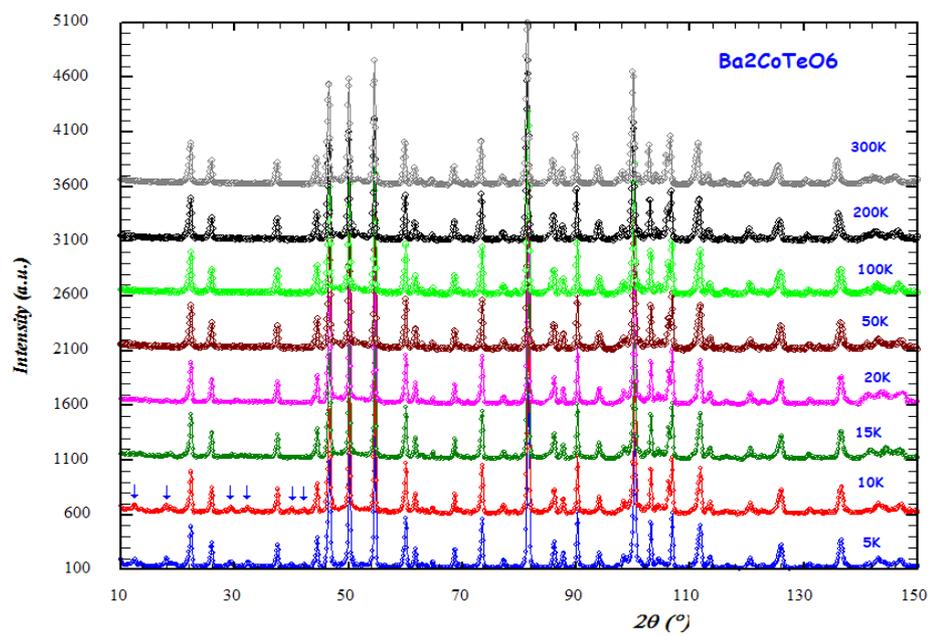

Figure 7

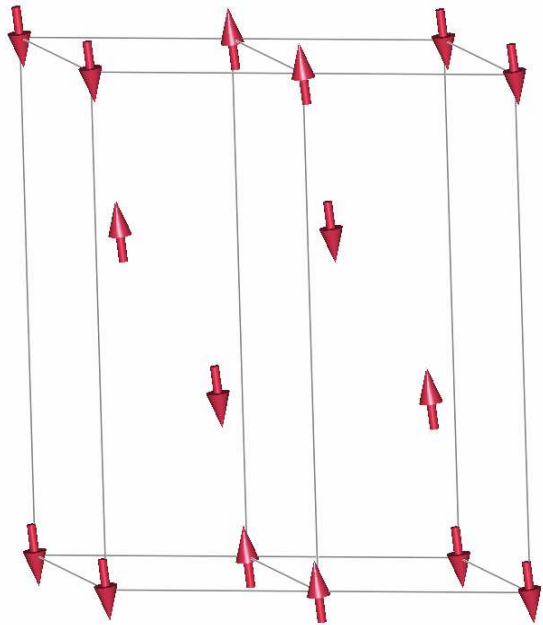

Figure 8

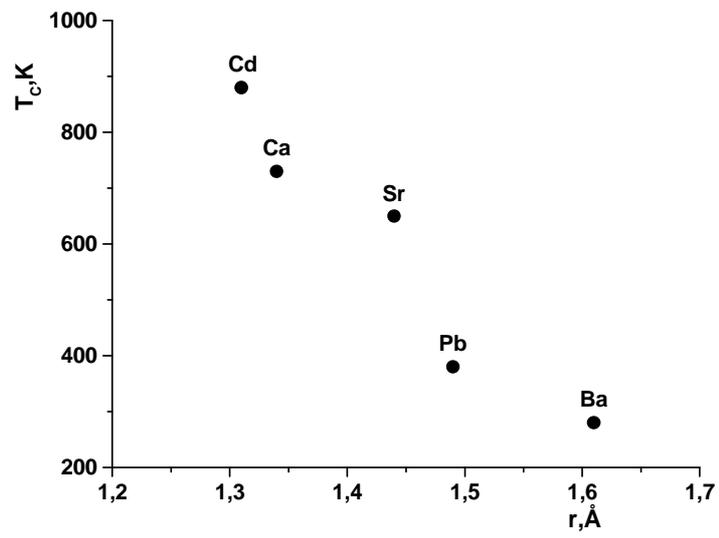

Figure 9